# IrisNet: Deep Learning for Automatic and Real-time Tongue Contour Tracking in Ultrasound Video Data using Peripheral Vision

M. Hamed Mozaffari [1*], Md. Aminur Rab Ratul [2], Won-Sook Lee [3]

[1-3] School of Electrical Engineering and Computer Science, University of Ottawa, 800 King-Edward Ave., Ottawa, Canada
[*]mmoza102@uottawa.com

**Abstract:** The progress of deep convolutional neural networks has been successfully exploited in various real-time computer vision tasks such as image classification and segmentation. Owing to the development of computational units, availability of digital datasets, and improved performance of deep learning models, fully automatic and accurate tracking of tongue contours in real-time ultrasound data became practical only in recent years. Recent studies have shown that the performance of deep learning techniques is significant in the tracking of ultrasound tongue contours in real-time applications such as pronunciation training using multimodal ultrasound-enhanced approaches. Due to the high correlation between ultrasound tongue datasets, it is feasible to have a general model that accomplishes automatic tongue tracking for almost all datasets. In this paper, we proposed a deep learning model comprises of a convolutional module mimicking the peripheral vision ability of the human eye to handle real-time, accurate, and fully automatic tongue contour tracking tasks, applicable for almost all primary ultrasound tongue datasets. Qualitative and quantitative assessment of IrisNet on different ultrasound tongue datasets and PASCAL VOC2012 revealed its outstanding generalization achievement in compare with similar techniques.

## 1. Introduction

Ultrasound technology, as a non-invasive and clinically safe imaging modality, has enabled us to observe human tongue movements in a real-time speech [1]. Analysis of the tongue gestures provides valuable information for many applications, including the study of speech disorders [2], Silent Speech Interfaces [3], tongue modelling [4] and second language pronunciation training [5] to name a few. By placing an ultrasound transducer under the chin of a subject in cross-sectional planes (i.e. mid-sagittal or coronal), researchers can see a noisy, white, and thick curve on display in real-time. In the mid-sagittal view, this curve is often presenting the place of the tongue surface (dorsum area). As a one-pixel diameter contour, it is the region of interest in different studies [6]. Red and yellow curves in Fig. 1 are segmented tongue surface and contour, respectively.

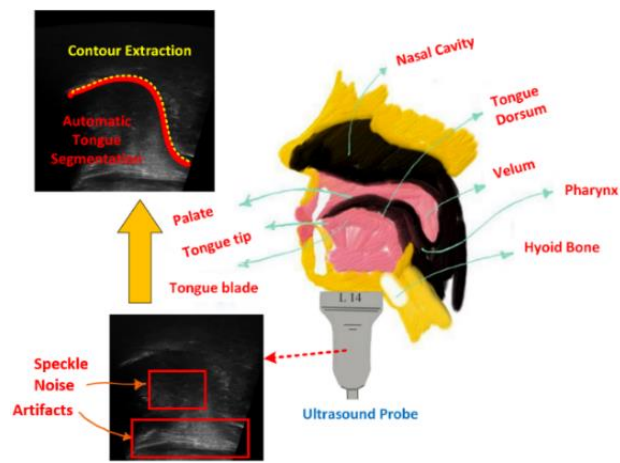

**Fig. 1.** Anatomy of the human tongue in mid-sagittal view. Tongue surface is relatively a sharp transient from dark to bright in ultrasound images.

A fully automatic and robust method is inevitable for tracking the tongue contour in real-time ultrasound video frames with a long duration. Images of tongue in ultrasound videos are quite noisy, accompany many high-contrast artefacts sometimes similar but unrelated to the tongue surface (see Fig. 1 for examples of speckle noise and shadow artefact). Despite significant efforts of researchers, a general, automatic, and real-time contour tracking method applicable for data recorded by different institutes using different ultrasound devices remains a challenging problem [7], [8]. Although a variety of image processing techniques have been investigated for tracking the tongue in ultrasound data automatically, manual re-initialization around the region of interest is still needed [8]. Gradient information of each ultrasound frame is required for many states of the art automatic tongue contour extraction methods [9]–[11]. Furthermore, pre- and post-processing manipulations are vital stages for almost all current techniques such as image cropping to keep the target region smaller and more relevant.

Recently, the versatility of deep convolutional neural networks has been proved in many computer vision applications with outstanding performance in tasks such as object detection, recognition, and tracking [12]. Medical image analysis, like image classification using deep learning, is a successful and flourishing research example [13], [14]. Medical image segmentation might be considered as a dense image classification task when the goal is to categorize every single pixel by a discrete or continuous label [13]. Depend on the definition of categories, image segmentation methods might be classified into semantic (delineate desired objects with one label) [15], [16] or instance (each specific desired object has a unique tag) [17].

Accessibility big datasets, accurate digital sensors, optimized deep learning models, efficient techniques for training neural networks, and faster computational facilities such as GPU have enabled researchers to implement reliable, real-time, and accurately robust systems for different applications. Deep learning techniques are currently used in many real-time applications from autonomous vehicles [18] to medical image intervention [19]. The recent investigation of machine learning techniques for the problem of tongue contour tracking revealed the outstanding performance of deep learning models for the challenge of automatic tongue contour tracking [7], [20]–[22]. Despite all achievements of new deep learning methods for tongue contour tracking, those methods suffer from artefacts in ultrasound data where initialization, enhancements, and manipulation are still required to alleviate false predictions. Furthermore, the generalization of the current deep learning model is not enough to use them for any ultrasound tongue dataset without fine-tuning [21].

As an alternative to fine-tuning, a specially designed deep learning model can be optimized well on small datasets such as tongue contours [23]. Fortunately, ultrasound video data from different institutions and machines contain similar noise patterns with limited artefact structures. In this study, our goal is to introduce a new deep convolutional neural network named IrisNet, proper for the task of real-time and automatic tongue contour tracking, benefiting from a new deep learning module named RetinaConv. Inspiring from peripheral vision property of the human eye, IrisNet can detect and delineate the region of interest faster and more accurate in comparison with recent techniques. One unique feature of IrisNet is its robust generalization ability, whereas it can be applied for novel ultrasound datasets without the use of fine-tuning and standard manual enhancements. To compare the performance of IrisNet with the state of the art models in the field of ultrasound medical image analysis [24], we assessed all models on the PASCAL VOC2012 dataset.

The remainder of this paper is structured as follows. A quick literature review of the field is covered in section 2. Section 3 describes our methodology in detail, including RetinaConv as well as the architecture of the IrisNet model. The experimental results and discussion around our proposed segmentation techniques for ultrasound tongue contour tracking comprise of system setup, dataset properties, and comparison studies are covered in section 4. In section 5, the performance of each model is tested on the PASCAL VOC2012 dataset. A real-time application of IrisNet is investigated in section 6. Section 7 concludes and outlines future work directions.

## 2. Literature review and related works

Exploiting and visualizing the dynamic nature of human speech recorded by ultrasound medical imaging modality provides valuable information for linguistics, and it is of great interest in many recent studies [10]. Ultrasound imaging has been utilized for tongue motion analysis in treatment of speech sound disorders [25], comparing healthy and impaired speech production [10], second language training and rehabilitation [26], Salient Speech Interfaces [3], and swallowing research [27], 3D tongue modelling [4], to name a few. Ultrasound data interpretation is a challenging task for non-expert users, and manual analysis of each frame suffers from several drawbacks such as bias depends on the skill of the user or quality of the data, fatigue due to a large number of image frames to be analysed, and lack achieving reproducible results [28].

Various methods have been utilized for the problem of automatic tongue extraction in the last recent years such as active contour models or snakes [8], [29]–[32], graph-based technique [33], machine learning-based methods [20], [34]–[37], and many more. A complete list of new tongue contour extraction techniques can be found in a study by Laporte et al. [10]. Although the methods above have been applied successfully on the ultrasound tongue contour extraction, still manual labelling and initialization are frequently needed. Users should manually label at least one frame with a restriction of drawing near to the tongue region [10], [38]. For instance, to use Autotrace, EdgeTrak, or TongueTrack software, users should annotate several points on at least one frame [38].

Moreover, due to the high capture rate and noisy characteristic of ultrasound device and rapidity of tongue motion, there might be frames with no valuable or visible tongue feature or with non-continues dorsum region. Therefore, for real-time tongue contour tracking, which users might move ultrasound transducer during exam sessions, a reliable and accurate technique is required to handle difficulties of real-time tasks applicable for other similar ultrasound tongue datasets [6].

Fully convolutional neural (FCN) networks were successfully exploited for the semantic segmentation problem in a study by [16]. Instead of utilizing a classifier in the last layer, a fully convolutional layer was used in FCN to provide a prediction map as the model's output [39]–[41] for any arbitrary input image size. Performance of FCN network models have been improved by employing several different operators such as deconvolution [42], concatenation from previous layers [43], [44], using indexed un-pooling [45], adding post-processing stages such as CRFs [46], employing dilated convolution [47], consider feature information from different stages in the model [48], utilizing region proposals before segmentation [49], and recently hierarchical structures [50].

Many of these innovations significantly improved the accuracy of segmentation results, relying on weights of a pre-trained model such as VGG16 [39] or DenseNet [51] which had been trained on a huge dataset such as ImageNet [52] or with the expense of more computational costs due to the significant number of network parameters. Therefore, due to the lack of a substantial general ultrasound dataset, state of the art techniques are not yet applicable for medical ultrasound image segmentation [23].

Few studies have been applied to deep semantic segmentation for the problem of ultrasound tongue extraction. In [53], Restricted Boltzmann Machine (RBM) was trained first on samples of ultrasound tongue images and ground truth labels in the form of encoder-decoder networks. Then, the trained decoder part of the RBM was tuned and utilized for the prediction of new instances in a translational fashion from the trained network to the test network. To automatically extract tongue contours without any manipulation on a large number of image frames, modified versions of UNet [43] have been used recently for ultrasound tongue extraction [6], [22], [54].

Similar to adding CRFs as a post-processing stage for acquiring better local delineation [46], more accurate segmentation results can be achieved when multiscale contextual reasoning from successive pooling and subsampling layers (global exploration) is combined with the full-resolution output from dilated convolutions [13], [48]. In this work, we proposed RetinaConv, a new convolutional module inspired by the peripheral vision of the human eye, utilizing the feature extraction ability of both dilated and standard convolutions. We evaluated our proposed architectures on a challenging dataset of ultrasound tongue images. The experimental results demonstrate that our fully-automatic proposed models are capable of achieving accurate predictions, with real-time performance, applicable for other ultrasound tongue datasets due to its strong generalization ability.

To investigate the performance of IrisNet in terms of accuracy, robustness, real-time, and generalization, we also conducted a real-time experiment in the field of Second Language (L2) pronunciation training. In this application, users can see their tongue in real-time superimposed on their faces. A fully automatic method should track the tongue surface in real-time during the user's speech. We also study the efficiency of IrisNet on data from other ultrasound machines in different institutions.

## 3. Methodology

Employing down-sampling techniques in deep learning models such as max-pooling layers has resulted in better contextual predictions in many major computer vision tasks such as image classification and detection [43]. In the field of image segmentation, a desirable result should contain accurately delineated regions with a contour around the target object. Due to the loss of information, down-sampling provides lower resolution prediction maps, which is not desirable for image segmentation tasks. Omitting pooling layers and replacing them with new operations such as dilated convolutions in the semantic segmentation field is a new idea. Unlike down-sampling, dilated convolution can keep the receptive field of a deep learning model [47].

In many recent publications, dilated convolution outperformed encoder-decoder techniques but with introducing grinding artefact to the results [55]. Recent studies showed that using feature maps from different layers of a network in the shape of encoder-decoder increases the performance of a model [48]. Therefore, benefiting from both encoder-decoder style and dilated architecture, novel models could find better image segmentation results [48]. Nevertheless, in the field of medical image analysis, UNET style architectures have still been popular and outperformed other techniques [14], [24], [56], [57]. The main reason for the significant achievements of top deep learning models in semantic segmentation is because of using pre-trained models on huge datasets. For the medical ultrasound image analysis field, there is still no such a massive and general dataset to use in the pre-training stage [23]. For this reason, the best alternative for current research progress is to optimize networks or using specific smaller models for each dataset [23].

Recently, specific deep learning networks have been utilized for the problem of ultrasound tongue contour tracking [6], [7]. The generalization ability of those methods is also investigated for other datasets [21]. However, fine-tuning is a vital step for using a model to work on a novel dataset. Note that negative transfer is another issue for using a pre-trained model for different datasets [21], [23]. In total, the generalization of a deep learning model for image segmentation is not clear, and the best method for a deep learning model is fine-tuning for only the imagery of a specific image context [58].

Fortunately, different ultrasound tongue datasets have similar characteristics such as point of view, which is often mid-sagittal cross-section, bright thick line in specific regions of the image (around 8 cm [30]), and almost similar image resolutions. Furthermore, possible gestures of the tongue are limited to the alphabet range of the speaker's language [59]. Moreover, speckle-noise patterns and artefacts are almost analogous for different subjects due to the limited movements of ultrasound transducer during data acquisition, relatively similar human oral region, and stable reference points such as palate or jaw hinge in datasets.

Therefore, we attempt to design a general and accurate deep learning model applicable for almost all standard ultrasound datasets, with the capability of real-time performance without any fine-tuning or image pre- or/and post- enhancements.

### 3.1. RetinaConv

The procedure of the correct segmentation task by the human brain has been unclear for researchers [58]. However, we know that the human eye has the ability of peripheral vision. One crucial strength of the human eye is to detect objects and movements outside of the direct line of sight, away from the centre of gaze. This ability, called peripheral (side or indirect) vision, helps us to detect and sense objects without turning our head or eyes, resulting in less computation for our brain.

Our vision around the central part of our eye's field of view is sharper than far from the centre. Following the human eye's peripheral vision ability, we designed a new convolutional module named RetinaConv. We simulated the idea of peripheral vision in the human eye by a convolutional filter module that is illustrated in Fig. 2, where the centre of the filter is stronger than around. One might consider this filter as a Gaussian filter as a combination of two kernels.

To make the RetinaConv filter module, we utilized the distributivity property of the convolution operator: $f * (g + h) = f * g + f * h$ where $f$ is the input image, $g$ and $h$ are standard and dilated convolutional filters, respectively. Therefore, applying two filters is equivalent to using the summation of them. By changing filter size and dilation factor, different peripheral vision strengths can be achieved for different sized images during the hyperparameter tuning stage. The benefit of RetinaConv is not limited to merely computational speed, but also to accuracy enhancement due to the use of two receptive fields simultaneously.

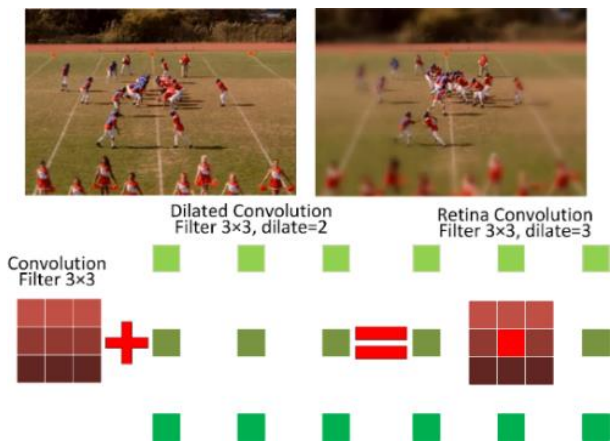

**Fig. 2.** Example of peripheral vision in the human eye. Centre of gaze is sharper due to more light detectors on Retina (dense kernel or standard convolution) and around is blurry because of fewer detectors on Retina (sparse kernel or dilated convolution).

### 3.2. Network Architecture (IrisNet)

Deep learning models such as DeepLab, with different versions (v1 [46], v2 [60], v3 [61], v3+ [48]) proposed by Google company, are the current robust networks which have been shown to be effective in semantic segmentation tasks for the context of natural images. Using DeepLab for ultrasound tongue imaging requires a pre-trained DenseNet model trained in advance on a huge source grey-scale ultrasound dataset (not in RGB format). A modified version of DeepLab v3 [61] without pre-trained weights was tested for tongue contour segmentation [7], the result was not significant. DeepLab models are also huge networks in terms of parameters that require a robust training system. On the other hand, UNET [43] is a ubiquitous model for medical image segmentation outperforming many other models [24] without the usage of pre-trained weights.

To maintain the powerful performance of deep learning models simultaneously, such as DeepLab and UNET, we designed the IrisNet network. As an encoder-decoder structure like UNET, IrisNet use extracted features from encoder layers passed to the decoder layers as well as max-pooling operator. However, the RetinaConv module in each layer keeps the receptive filed wider by utilizing dilated convolution. Individual kernels of RetinaConv highlights mid-point regions more by emphasizing the surrounding area. As a combination of kernels, grinding [55] and checkerboard artefacts [62] due to the inappropriate filter sambling rate are alleviated significantly.

However, analogous to a Gaussian filter, RetinaConv provides more blurry features. To address this issue, we used a dilation factor of 1 on both end sides of the network. The network architecture of IrisNet is presented in Fig. 3. The figure illustrates the RetinaConv module, encoder, and decoder blocks, as well. Using different operators such as max pooling, transpose convolution, skip connections and RetinaConv, IrisNet able to predict delineated regions by using both low- and high-level features at the same time results in better segmentation output.

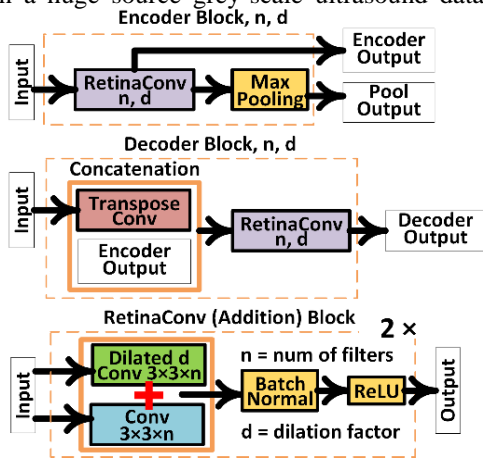
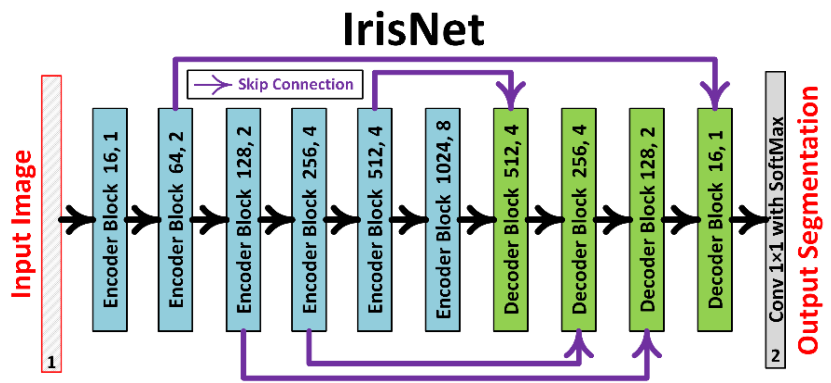

**Fig. 3.** Network Architecture of IrisNet. RetinaConv module is used in both encoder and decoder layers each for two times. Plus symbol is sum operator, while squared box around two kernels indicate concatenation operator.

Each encoder and decoder block in IrisNet has two times repeated RetinaConv blocks following by batch normalization (BN), ReLU activation function, and max-pooling layers. The number of filters and dilated factors are indicated in Fig. 3. Within each decoder block, feature maps from the previous layer first are up-sampled using 2×2 transposed convolution. Then the output is concatenated by the result of the corresponding encoder block skipped from the later encoder section. In the last layer, we used a 1×1, fully convolutional layer.

In former deep learning models, which have been proposed for automatic tongue contour tracking [20]–[22], the sigmoid activation function in the output layer provides on channel grey-scale instances. Therefore, ground truth labels should be grey-scale with 255 classes after data augmentation. Following semantic segmentation literature, we developed the IrisNet model to predict instances comprises of two binary channels, one for foreground and another for background labels. Therefore, the last layer activation function in IrisNet is SoftMax, and ground truth labels are in a binary format even after data augmentation.

Although this method of using two classes is not a new idea for the medical image segmentation field [43], it is a novel idea for the ultrasound tongue contour tracking. The importance and impact of using background information in the network training stage are that the model can also learn which area of an image belongs to the background. See Fig. 1 for an example of artefacts that can be detected as a tongue surface but its artefact shadow from the jaw. Therefore, IrisNet can easier discriminate background artefacts and noise from the region of interest. Furthermore, this technique releases researchers from cropping extra information such as ultrasound settings, dataset brand, or palate/jaw regions. Our experimental results also revealed the importance of using two classes instead of one in this literature.

## 4. Experimental Results

We evaluated the performance of IrisNet for automatic tongue contour tracking tasks in ultrasound image sequences after levels of hyperparameter tuning. We also investigated the performance of IrisNet on the PASCAL VOC2012 dataset.

### *4.1. Dataset*

There are plenty of ultrasound datasets private or publicly available for training deep learning models for automatic tongue contour tracking such as seeing speech project (SSP) [59], University of Michigan (UM) [22], and University of British Columbia (UBC) [63], to name the most common once. However, none of them provides annotated data. To test the IrisNet model, we utilized ultrasound data from the University of Ottawa (UO) comprises of 2085 annotated masks by two experts [7]. For each image, there are two ground truth labels, one for background and another for the foreground (See **Fig. 4** for more details).

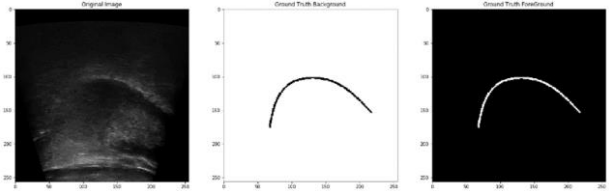

**Fig. 4.** A sample frame from the UO dataset. The middle and right images are a background and foreground truth labels (white and black are one and zero), respectively.

We divided the dataset into training, validation, and test sets by 80%, 10%, and 10% ratios, respectively. To increase dataset size, we used online data augmentation with realistic transformation factors, common in ultrasound datasets of the tongue, including horizontal flipping, rotation by a maximum range of 25 degrees, translation of 40 pixels shift in each direction and zooming from 0.5x to 1.5x scale. It is noteworthy that ground truth data in semantic segmentation literature are in one hot encoding format (binary images). Nevertheless, in the tongue contour tracking field, many previous studies used grey-scale labels for the training of their machine learning models [20], [22]. In this study, we followed the method in [7] for binarization of ground truth and online data annotation. For this reason, ground truth labels are gagged after binarization and even in the testing stage.

Following the method in [64], we calculated the diversity of several datasets with respect to the average of each dataset. For instance, a comparison between UO and SSP datasets is depicted in **Fig. 5**, where UO had the most heterogeneous data in contrast to other datasets. Accordingly, UO might be considered as a challenging ultrasound tongue dataset.

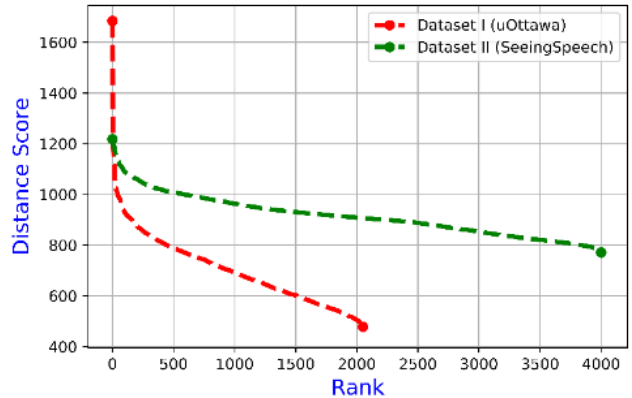

**Fig. 5.** Diversity of datasets in terms of distance from their average image.

### *4.2. Training and Validation*

We trained IrisNet by Adam optimization [65] method using its default parameters 0.9 and 0.99 for $\beta_1$ and $\beta_2$, respectively. We use a minibatch size of 20 images for 50 epochs. We test the sensitivity of the learning rate for IrisNet with different decay factors using a variable learning rate. The result can be better with different learning rates. For the sake of a fair comparison between models, we used a fixed learning rate of $10^{-3}$ for IrisNet, chosen by line search, and for other models, we utilized their default values from each publication. For all models, we used random initialization for network weights. We implemented and tested all models using one NVIDIA 1080 GPU unit, which was installed on a Windows PC with Core i7, 4.2 GHz speed, and 32 GB of RAM. We also used the Google Colab with a Tesla K80 GPU and 25GB of memory for acquiring training results faster in parallel.

To validate each model, we used a validation set for each dataset. Dice loss and Binary Cross-Entropy loss [7] were utilized for validation criteria. The training trend of IrisNet is presented in **Fig. 6**. In both diagrams, satisfactory progress can be observed in terms of over-fitting and under-fitting. Less fluctuation was seen in our experiments for IrisNet in compare to other methods of this work. Note that Dice loss is (1 – Dice Coefficient), and it should be minimized.

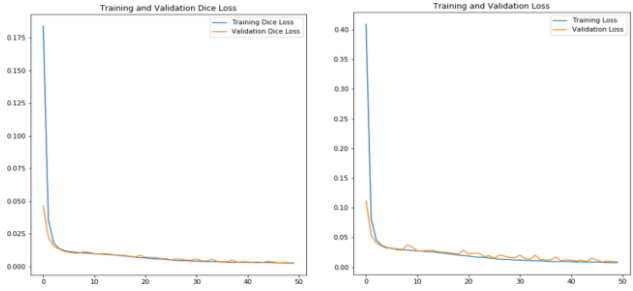

**Fig. 6.** Training trend of IrisNet model on Ultrasound data. Left to right: Dice Loss and Binary Cross Entropy.

### 4.3. Qualitative Study

To assess IrisNet for the task of tongue contour tracking in ultrasound data, we train and compare results of recent and original similar deep learning models in the literature, including UNET [43], FCN8 [16], BowNet and wBowNet [7]. For each model, we used default values from each publication for their parameters. System setting, training procedure, annotated data, random seeds, and validation method all were selected similarly for each model. We used checkpoint saving instead of stop criteria for the training of each model to keep the best-trained models. **Fig. 8** presents our qualitative results of IrisNet for five randomly selected frames from the test set. IrisNet predicts less noise and false prediction in comparison to the other models. FCN8 predictions contain grey-scale squared shape because of using up-sampling instead of transpose convolution in the decoder section.

### 4.4. Quantitative Study

To see which model predicted instances with less significant noise, we use the same threshold value for all models as 0.1. **Table 1** presents values of intersection over union [46] before and after thresholding instances as well as the average value for all samples in the test set (mIOU). IrisNet could predict instances with less noise than other models. Note that the same test images have been used in both qualitative and quantitative evaluations.

**Table 1.** Intersection Over Union (IOU) values before and after thresholding. The same threshold value and test images were used for all models (tIOU).

| Image | UNET | | BowNet | | wBowNet | | IrisNet | | FCN8 | |
|---|---|---|---|---|---|---|---|---|---|---|
| | IOU | tIOU | IOU | tIOU | IOU | tIOU | IOU | tIOU | IOU | tIOU |
| (1) | 87.7 | 38.9 | 84.1 | 36.8 | 87.7 | 40.7 | **87.8** | **40.8** | 87.7 | 32.8 |
| (2) | 91.7 | 42.5 | 89.2 | 38.4 | 91.7 | 42.8 | **91.7** | **42.9** | 91.7 | 37.1 |
| (3) | 94.6 | 48.4 | 94.3 | 54.3 | 94.6 | 49.5 | **94.6** | **53.2** | 94.6 | 39.7 |
| (4) | 88.9 | 39.5 | 87.1 | 33.0 | 88.9 | 36.5 | **88.9** | **36.6** | 88.9 | 35.9 |
| (5) | 83.4 | 39.2 | 81.1 | 42.1 | 83.4 | 40.4 | **83.5** | **45.3** | 83.4 | 38.7 |
| mIOU | 98.3 | 51.2 | 98.1 | 87.5 | 98.1 | 50.0 | **98.4** | **51.5** | 97.8 | 50.0 |

### 4.5. Linguistics Criteria

There are many linguistics methods for the evaluation of tongue contour tracking accuracy. However, we employed the standard techniques in the literature for testing machine learning models. To extract contours from predicted segmentation results, first, we apply a skeletonization method on thresholded predictions (see **Fig. 9**) and ground truth labels following the method in [9] (see **Fig. 7**). As can be seen from **Fig. 8** and **Fig. 9**, IrisNet provides better instances in almost all cases in terms of disconnected regions and noisy segments in contrast to other similar techniques.

For the same experiments, values of Mean Sum of Distances in terms of pixel and millimetres are reported in **Table 2**, while IrisNet could provide MSD values with better accuracy. We investigated the generalization ability of IrisNet for other datasets. For this reason, we selected random frames from each common publicly available dataset (EdgeTrak [30], UBC [63], SSP [59], Ultrax [66], UM [22], UA [67]) and test IrisNet on each of those datasets.

**Table 2.** Mean and standard deviation of Mean Sum of Distances (in pixels, 1 pixel ≈ 0.15mm) for 280 frame test datasets.

| Model | MSD (px) | MSD (mm) |
|---|---|---|
| UNET | 4.15±0.72 | 0.62±0.39 |
| BowNet | 4.58±0.39 | 0.69±0.54 |
| wBowNet | 4.38±0.37 | 0.66±0.07 |
| IrisNet | **4.12±0.26** | **0.61±0.12** |
| FCN8 | 5.05±0.15 | 0.76±0.08 |

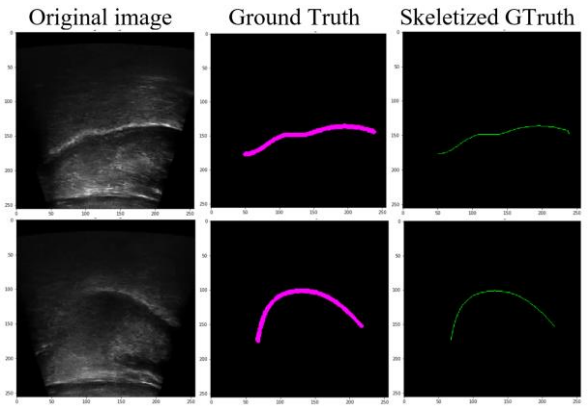

**Fig. 7.** The curve (green) is skeletonized results determined from the ground truth label (pink).

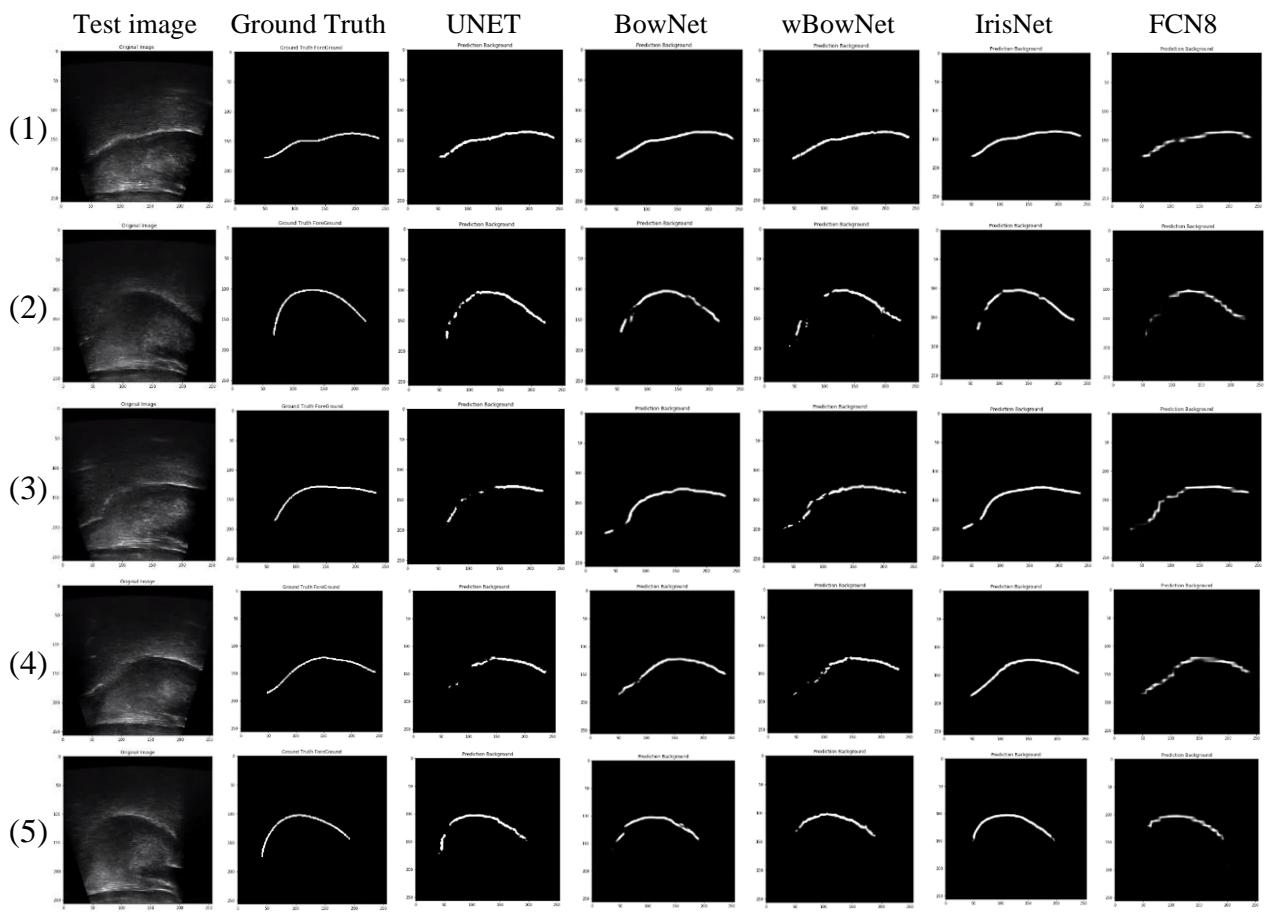

**Fig. 8.** Sample results from the qualitative study. Each column shows the foreground predicted by each deep learning model. No pre- or post-processing has been applied for the dataset.

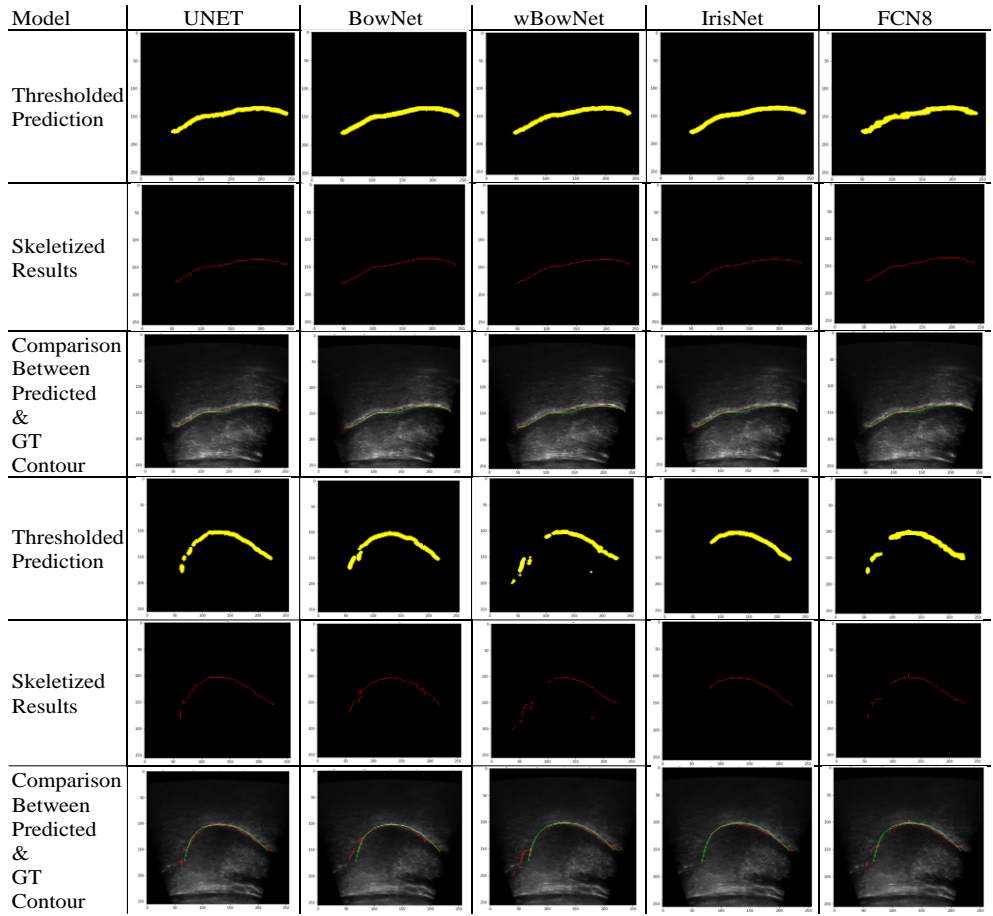

**Fig. 9.** Comparing the contour extracted from segmentation results (red curves) with the contour obtained from ground truth labels (green curves) of each deep learning model for the two test samples of the previous figure.

From Fig. 10, although IrisNet had never seen any sample from testing datasets, it could predict segmentation results without any issue. Besides the generalization ability of IrisNet, one reason for this ability is that the test datasets from other institutes is relatively simple but with similar feature to the source UO dataset. Note that for the sake of representation, we warp test datasets. The qualitative results of the same study can be seen in **Table 3**. Except for the UA dataset, for almost all other datasets, IrisNet could predict better instances on average in terms of MSD.

**Table 3.** Mean and standard Deviation of Mean Sum of Distances (in Pixels) for different test datasets comprises of 20 randomly selected frames.

| Model | UM [22] | SSP [59] | UBC [63] | UA [67] |
|---|---|---|---|---|
| UNET | 5.27±0.81 | 6.31±0.25 | 5.42±0.73 | **6.83±0.53** |
| BowNet | 5.41±0.26 | 7.83±0.74 | 6.73±0.23 | 8.35±0.26 |
| wBowNet | 5.38±0.97 | 6.63±0.26 | 5.83±0.64 | 7.74±0.59 |
| IrisNet | **5.29±0.10** | **6.27±0.85** | **5.15±0.73** | 6.87±0.48 |
| FCN8 | 6.53±0.73 | 7.73±0.98 | 6.62±0.12 | 8.73±0.78 |

### *4.6. Performance comparison*

Although IrisNet is superior to state-of-the-art deep learning models in ultrasound tongue segmentation literature, it has more trainable parameters. On the other hand, it has faster performance in real-time applications due to the efficient implementation of the model. In **Table 4**, IrisNet has more parameters than the original UNET, but it has similar performance speed using GPU.

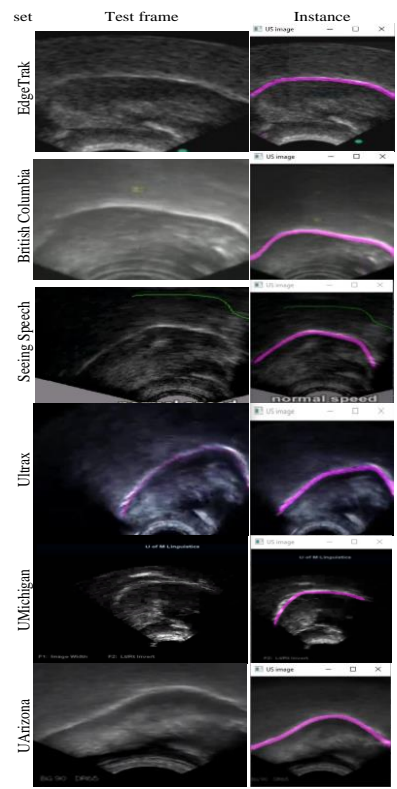

**Fig. 10.** Testing IrisNet on sample data from standard ultrasound tongue image datasets.

**Table 4.** Performance speed (FRate in frames per second) and the number of trainable parameters (Params in millions). All models tested on GPU for ten times. The average and standard deviation values are reported for FRate.

|  | UNET | BowNet | wBowNet | IrisNet | FCN8 |
|---|---|---|---|---|---|
| Params | 31.1m | 0.92m | **0.79m** | 5.93m | 134.2m |
| FRate | 45±0.25 | 43±0.14 | 32±0.25 | **44±0.83** | 27±0.72 |

## 5. Evaluation of PASCAL VOC2012 Dataset

As we mentioned in section 3, state of the art semantic segmentation techniques are powerful for natural image datasets such as PASCAL VOC 2012 [68] due to their pre-trained blocks, trained on massive datasets. For this reason, we investigate the ability of the most common deep learning model for ultrasound image segmentation literature, UNET [43] with and without the use of pre-trained weights. We also tested IrisNet, BowNet models [7], and pre-trained FCN8 on Pascal Dataset.

The PASCAL dataset has 2621 images of 21 classes. We divided the dataset into train/validation/test sets with a ratio of 80/10/10 percentages. We also applied online augmentation during the training of each model, including horizontal flipping, 20 pixels shift range in each direction, normalizing from 255, and rotation by the maximum of 5 degrees. Each model trained on the dataset for 50 epochs and the best model was saved for testing.

Adam optimization with default parameter values of 0.9 and 0.999 for $\beta_1$ and $\beta_2$, respectively, was used to optimize each model using Binary Cross-Entropy Loss and F1 score loss. IOU and mIOU were reported in the testing stage for all test set samples. The learning rate was variable for all models starts from $10^{-6}$ and decay factor of 0.001. We used Google Colab with a Tesla T4 GPU for training and testing of models. To find the best deep learning models, we also tested Sigmoid and SoftMax activation functions for each model separately. As can be seen from Fig. 11 and Table 5, UNET model performance can be improved as a powerful model when the model is fine-tuned using VGG16 encoder blocks pre-trained on the ImageNet dataset.

On the other hand, from Fig. 12, average IOU for IrisNet is like fine-tuned versions of UNET and FCN8. Our experimental results showed that for the PASCAL dataset, without using weights of a pre-trained model such as VGG16, the FCN8 model provides empty instances, and UNET instances are considerably weak (see Fig. 11 predictions of dog and cat from original UNET). For fine-tuned versions of UNET, performance is significantly better. Domain adaptation for ultrasound tongue contour tracking was investigated recently [21] with similar consequences. It is noteworthy that IrisNet could predict acceptable segmentation instances in compare to fine-tuned versions of FCN8 (with two times more trainable parameters) due to the better generalization ability of IrisNet.

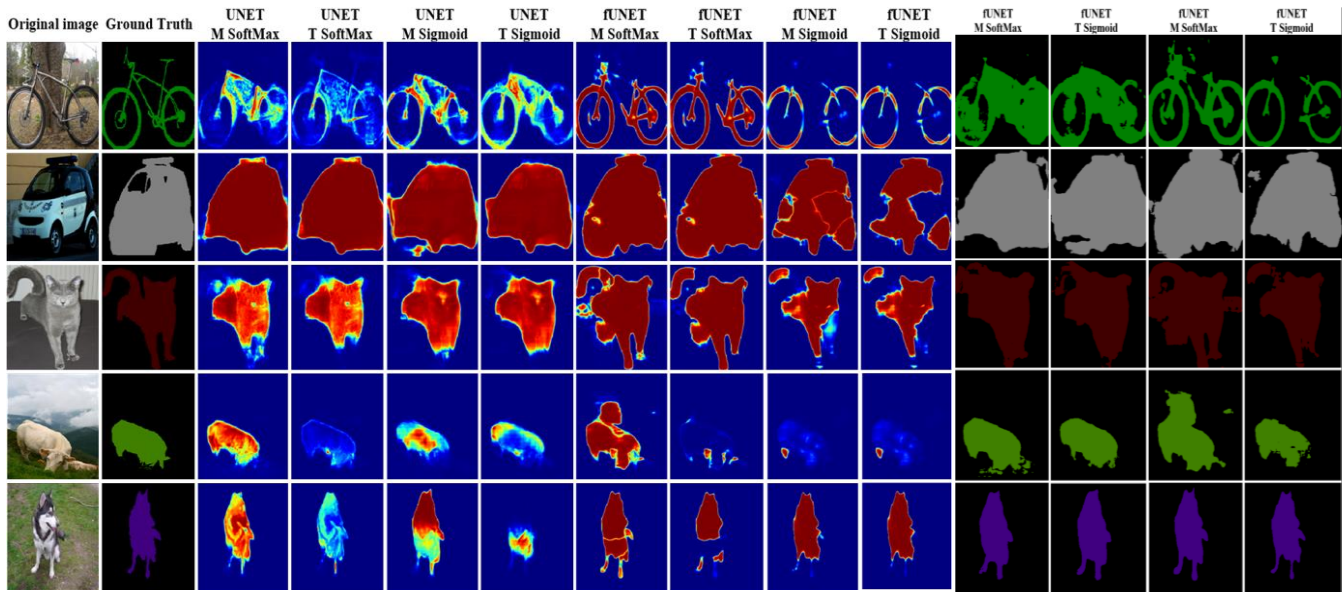

**Fig. 11.** Sample results of the UNET model with different settings. M and T indicate the average results on all output channels and target channel, respectively. UNET is the original model trained from scratch, while fUNET is fine-tuned on the VGG16 model pre-trained on the ImageNet dataset. From left to right, columns 3 to 10 are feature maps. Columns 11 to 14 are thresholded results.

**Table 5.** IOU values for the UNET model with and without fine-tuning over ImageNet dataset. SM and SI are SoftMax and Sigmoid activation function in the last layer of UNET. fUNET is a fine-tuned version.

| Model | Trainable Params (million) | Airplane | Bicycle | Bird | Boat | Bottle | Bus | Car | Cat | Chair | Cow | Table | Dog | Horse | Motorbike | Person | Plant | Sheep | Sofa | Train | TV | mIOU |
|---|---|---|---|---|---|---|---|---|---|---|---|---|---|---|---|---|---|---|---|---|---|---|
| U-net SM | 31.1 | 60.3 | 32.1 | 54.6 | 10.9 | 41.4 | 75.5 | 74.6 | 74.1 | 21.8 | 70.4 | 43.7 | 85.1 | 69.5 | 55.4 | 77.3 | 28.8 | 73.6 | 53.6 | 16.3 | 45.2 | 58.9 |
| U-net SI | 31.1 | 61.8 | 39.6 | 56.8 | 27.3 | 53.5 | 74.6 | 78.6 | 73.8 | 18.3 | 74.9 | 10.2 | 84.3 | 65.3 | 53.0 | 79.0 | 07.4 | 72.5 | 70.6 | 31.4 | 17.8 | 55.7 |
| fU-net SM | 11.1 | **68.3** | 51.9 | 53.5 | 18.0 | 41.6 | **82.3** | 81.4 | 62.1 | 19.8 | 80.5 | **49.7** | 85.3 | **70.8** | 85.5 | 61.4 | 25.5 | 74.6 | **82.3** | 35.3 | **61.5** | **66.4** |
| fU-net SI | 11.1 | 46.9 | **60.4** | **64.3** | **34.3** | **48.3** | 81.6 | **82.4** | **85.8** | **51.4** | 81.3 | 44.4 | **93.3** | 51.6 | **89.3** | **85.7** | **31.8** | **75.5** | 67.5 | **43.6** | 57.8 | 65.2 |

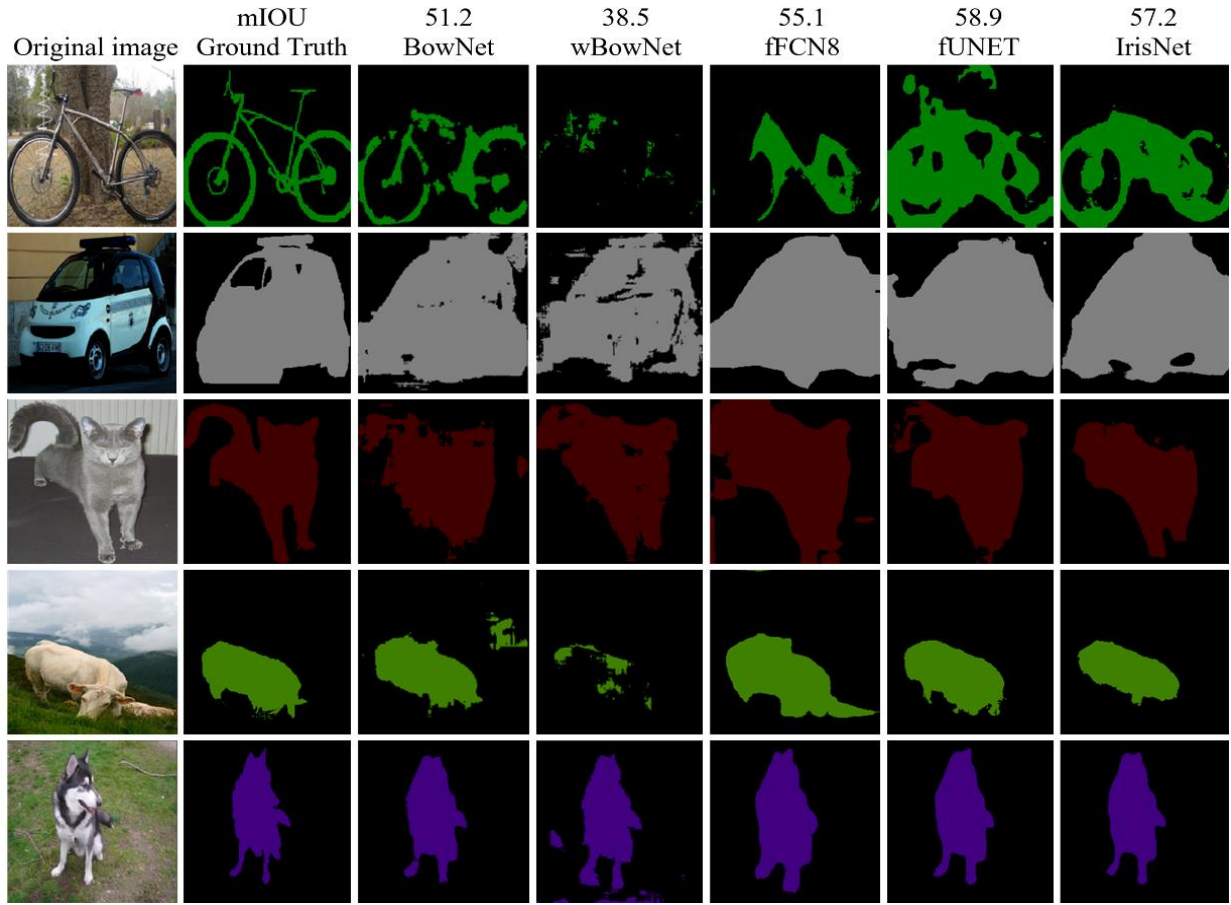

**Fig. 12.** Sample results of IrisNet, BowNet, FCN8, and UNET models. FCN8 and UNET were trained on pre-trained VGG16 weights. Average IOU is reported for each model on the test dataset.

## 6. IrisNet for Ultrasound-Enhanced Multimodal L2 Pronunciation Training System

Pronunciation is an essential aspect of Second Language (L2) acquisition as well as the first judgmental representation of an L2 learner's linguistic ability to a listener [69]. Teaching and learning of pronunciation of a new word for an L2 learner have often been a challenging task in traditional classroom settings with listening and repeating method [70]. Computer-assisted pronunciation training (CAPT) as biofeedback can alleviate this difficulty by visualization of articulatory movements [71].

In a separate study, we implemented a comprehensive system including several deep learning modules working in parallel for automatic and real-time visualization and tracking of ultrasound tongue as well as superimposing the result on the face side of L2 learners. Offline versions in previously proposed systems can be found in the literature [6], [72], [73]. For the sake of this study, we only report the tongue contour tracking module using IrisNet.

One screenshot from our system can be seen in Fig. 13. IrisNet works in real-time with high accuracy in parallel with other modules of our system. From the result of our experiments, using deep learning models such as BowNet or UNET, there will be a trade-off between high accuracy and real-time performance. IrisNet was also tested on datasets from UM in our real-time system (just ultrasound data was offline). The qualitative results were significantly better due to the higher resolution of UM dataset video data as well as less noise and artefacts in that dataset [22].

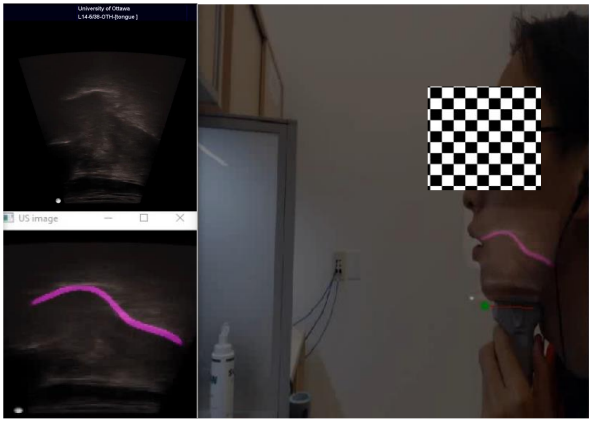

**Fig. 13.** Screenshot from our pronunciation training system. IrisNet automatically tracks tongue contour in real-time. Calibration data are determined by another deep learning model to superimpose ultrasound data on the user's face.

## 7. Conclusion and Future Work

In this paper, a novel deep learning model (IrisNet) is developed for segmentation and tracking of tongue contours in ultrasound video data. In the proposed algorithm, a new convolutional module (RetinaConv) is introduced, implemented, and tested, inspiring by human peripheral vision ability. RetinaConv module alleviates the difficulties of some issues due to the use of transpose and dilated convolutions for semantic segmentation models, including checkerboard and gripping artefacts. We demonstrate that the IrisNet model performs well compared with similar techniques for the task of automatic tongue contour tracking in real-time.

Following semantic segmentation literature, we considered background information as a separate class label in the training and validation process of our models. The consequence was the significant improvement of all deep learning models in the ability of discrimination between background artefacts and the region of interest, which is tongue contour. Our experimental results on a challenging dataset of tongue ultrasound illustrate the powerfulness of the IrisNet model in generalization. IrisNet can even predict excellent instances for a novel ultrasound tongue dataset without any fine-tunning or training.

Using deep learning models pre-trained on a huge source dataset such as ImageNet will result in better instances after fine-tuning on a target domain with the same context. However, fine-tuning is a difficult task with its disadvantages, such as negative training. Furthermore, a huge general dataset in ultrasound medical image analysis is not available yet. Therefore, IrisNet can be a promising alternative for the current small specific datasets as a general deep learning model for ultrasound image segmentation. We investigated the powerfulness of IrisNet on the PASCAL VOC2012 dataset, and its performance is acceptable in comparison with fine-tuned models with a huge embedded pre-trained model such as VGG16. To show the generalization performance of IrisNet, we employed that for tracking tongue contour on other tongue ultrasound datasets as well as in a real application in second language pronunciation training with significant results in terms of generalization, accuracy, and real-time performance.

Although IrisNet and its proposed RetinaConv module outperform existing deep models in the literature, performance evaluation of the model on other datasets is a question. Furthermore, performance can be improved by using variable learning rates and increasing training dataset size by combining several sets. Currently, the research team is evaluating the ability of IrisNet to track the whole tongue region instead of the tongue contour.

We believe that publishing our datasets, annotation package, and our proposed deep learning architectures, all implemented in multiplatform python language with an easy to use documentation can help other researchers in this field to fill the gap of using previous methods where several non-accessible requirements are needed as well as they customized for restricted datasets. Evaluation of IrisNet for other medical ultrasound image segmentation tasks can be a future of this work.